\documentclass[sort&compress]{elsarticle}
\usepackage{graphicx}
\newcommand{\chem}[1]{\ensuremath{\mathrm{#1}}}
\newcommand{\un}[1]{\ensuremath{\unskip\,\mathrm{#1}}}

\begin{document}

\title{In-phase and anti-phase synchronization in noisy
Hodgkin-Huxley neurons}

\author{Xue Ao}

\author{Peter H\"anggi}

\author{Gerhard Schmid}
\ead{Gerhard.Schmid@physik.uni-augsburg.de}
\address{Institut f\"ur Physik, Universit\"atsstr. 1, 86159 Augsburg, Germany}

\begin{abstract}
We numerically investigate the influence of intrinsic channel noise
on the dynamical response of delay-coupling in neuronal systems.
The stochastic dynamics of the spiking is modeled within a
stochastic modification of the standard Hodgkin-Huxley model wherein the delay-coupling accounts for the
finite propagation time of an action potential along the neuronal axon. We quantify this  delay-coupling
of the Pyragas-type in terms of the difference between corresponding presynaptic and postsynaptic membrane potentials.
For an elementary neuronal network consisting of two coupled neurons we
detect  characteristic stochastic synchronization patterns which exhibit
multiple phase-flip bifurcations: The phase-flip bifurcations occur in form of  alternate transitions
from an in-phase spiking activity towards an anti-phase spiking activity.
Interestingly, these phase-flips remain robust in strong channel noise and in turn cause a striking stabilization
of the spiking frequency.
\end{abstract}

\maketitle

\section{Introduction}
\label{sec:Intro}

Time-delayed feedback presents a common mechanism which is found in many biological systems including
neuronal systems. Signal transmission time delays in neuronal systems
either result from (i) chemical processes in the neuronal synapses where
neurotransmitters are released and diffusively overcome the synaptic cleft and/or (ii) from the finite propagation speed of
electrical excitations along the neuronal axon. Time delays stemming from chemical synapses are of the order of a
few milliseconds, while the axonal conduction delays in both, delay-coupled neurons
and autaptic feedback loops, reach values up to tens of milliseconds \cite{Masoller2008,Kandelbook,Loos1972,Lubke1996}.

As the time scale of the delayed coupling and of the neuronal dynamics become comparable, the delay-coupling
gives raise to peculiar synchronization phenomena \cite{Pikovsky2003}. In particular, phase synchronization phenomena
in neuronal systems are commonly thought to be the basis for many biological relevant processes occurring in the brain \cite{Varela2001, Lachaux1999}.
Synchrony of neurons from small brain regions up to large-scale networks of different  cortices comes along with
transmission time delays. Theoretical and computational studies on neuronal networks with delay-coupling
recently highlighted
the occurrence of so-called phase-flip bifurcations \cite{Prasad2006, Prasad2008, Adhikari2011}.
The ensemble activity of the coupled neurons change abruptly from  in-phase to anti-phase oscillations or vice versa.

With this work we research this objective by considering the influence of internal noise.
It is an established fact that noise leads to various prominent effects in neuronal dynamics \cite{Lindner2004}. Some typical examples that come to mind are stochastic resonance features
\cite{Gammaitoni1998, Hanggi2002, Schmid2001EPL, Jung2001EPL},
and noise-assisted  synchronization \cite{Pikovsky2003, Lacasta2002, Callenbach2002, Freund2003}.
Within our work the intrinsic noise is due to the stochastic gating of the ion channels,
i.e. the so-called channel noise which is inherently coupled to the electrical properties of the axonal
cell membrane \cite{White2000, Goldwyn2011, Wainrib2012}. Interestingly, it has been shown  that intrinsic channel noise
does not only lead to the generation of spontaneous action potentials \cite{Chow1996}, but as well affects the
neuronal dynamics at different levels,  namely: (i) it can boost the signal quality \cite{Schmid2001EPL, Jung2001EPL}, (ii)
enhance the signal transmission reliability \cite{Ochab2009},  (iii)  cause  frequency- and phase-synchronization features
\cite{Schmid2002,Casado2003PLA, Casado2003PRE, Hauschildt2006, Yu2007} and (iv) may result in a frequency
stabilization \cite{Li2010}, to name but a few.

The present work is organized as follows: In Sec.~\ref{sec:model} we introduce the biophysical model.
We review the standard Hodgkin-Huxley model and its generalizations with respect to intrinsic
channel noise and a delay-coupling. Numerical methods for simulation are introduced after that.
In Sec.~\ref{sec:two-neuron-network}, the dynamics of a network of two delay-coupled Hodgkin-Huxley neurons is explored both, in the  deterministic limit and under consideration of channel noise.
As a comparison, we retrospect on the previous work on a single neuron subjected to a delayed feedback loop resulting from autapse in Sec.~\ref{sec:retrospect}. Our  conclusions  are given in Sec.~\ref{sec:conclusion}.

\section{Biophysical model setup}
\label{sec:model}

We consider a minimal building block of a neuronal network composed of two coupled neurons. As an arche\-type model for nerve excitation of the individual neuron, we utilize a stochastic
generalization of the common Hodgkin-Huxley model.
The stochastic generalization accounts for  intrinsic membranal conductance fluctuations,
i.e. channel noise, being  caused by  random ion channel gating. Moreover, we account for a delay in the coupling which
accounts for a finite propagation time of the action potential along the axon.

\subsection{Hodgkin-Huxley-type modelling of two delay-coupled neurons}

According to Hodgkin and Huxley, the dynamics of the membrane potential $V_i$
with $i=1,2$ of  two coupled neuronal cells is given by \cite{hh}
\begin{eqnarray}
  \label{eq:voltage_equation}
  &C \frac{\mathrm{d}}{\mathrm{d}t} V_i  + G_\mathrm{K}(n) \, (V_i - V_\mathrm{K}) \nonumber \\
  & + G_\mathrm{Na}(m,h) \, (V_i - V_\mathrm{Na})
   + G_\mathrm{L} \, (V_i-V_\mathrm{L}) = I_{i}(t)\, .
\end{eqnarray}
Here, $V_i$ denotes the membrane potential of the $i$-th cell.
The stimulus $I_{i}(t)$ acting on the $i$-th neuron reads:
\begin{equation}
 \label{eq:currentstimulus}
 I_{i}(t) = I_{i \mathrm{, ext}}(t) + I_{i,j}^{\tau}(t)\,, \quad i,j = 1,2\, , i\neq j\, ,
\end{equation}
where the bi-directional delay-coupling of Pyragas-type \cite{Pyragas1992}
between the two neurons is assumed to be linear in the difference of the membrane potentials
of a primary, $i$-th neuron at time $t$
and a secondary, $j$-th neuron at an earlier time, $t-\tau$. The coupling thus reads:
\begin{equation}
 \label{eq:coupling}
 I_{i, j}^{\tau}(t ) = \kappa \, \left[ V_j(t-\tau) - V_i(t) \right]\, ,
\end{equation}
where $\kappa$ corresponds to the coupling strength and $\tau$ denotes the finite delay time.
The coupling defined in Eq.~(\ref{eq:coupling}) is of ``electrotonic'' type,
i.e. we consider an idealized situation wherein the coupling is proportional to the difference of
presynaptic and postsynaptic membrane potentials. This kind of coupling then  corresponds to so-called
gap-junctions which allow the bi-directional transport of ions and small molecules from one neuronal
cell into another. Unlike the conductance of chemical synapses, the conductance of gap-junctions is
independent of the presynaptic and postsynaptic membrane potential and can therefore be modelled by
the constant coupling parameter $\kappa$.
Possible chemical mechanisms occurring at the synaptic cleft
are assumed to be instantaneous as the time scale for signal propagation along the neurons axon is
much larger than the corresponding one for the transport process in the synaptic cleft.
Note, that the delayed  stimulus in Eq.~(\ref{eq:coupling}) results in an excitatory coupling mechanism
in which the spiking of neuron $i$ at an earlier time $t-\tau$ time
favors the initiation of a action potential of the other cell at time $t$.

In addition to the delayed, bilinear coupling current we apply
a constant current stimulus $I_{i, \mathrm{ext}}$ on the neurons, mimicking the common stimulus of the
neuronal environment on the so considered two-neuron network. In absence of the bi-directional coupling the dynamics
of each neuron exhibits a bifurcation scenario exhibiting a subcritical Hopf bifurcation.
As a consequence, the membranal dynamics displays
(i) a stable fix-point, i.e. the so-called {\itshape rest state}
for $I_{i, \mathrm{ext}} < I_1 \approx 6.26 \mu \mathrm{A/cm}^2$, (ii) a stable spiking solution for
$I_{i, \mathrm{ext}} > I_2 \approx 9.763 \mu \mathrm{A/cm}^2$ and (iii) a bistable regime for which the stable rest state and a stable
oscillatory spiking solution coexist, i.e. for $I_1<I_{i, \mathrm{ext}}<I_2$ \cite{Nelsonbook,Hassard1978,Rinzel1980,Schmid2003LNP,Izhikevich2000}.
In particular, for $I_{i, \mathrm{ext}}=0$ the membrane potential is $V_{\mathrm{rest}} = -65.0 \mathrm{mV}$.

Throughout this work the membrane potentials are measured in units of $\mathrm{mV}$
and time in units of $\mathrm{ms}$.
For a squid giant axon, the parameters in Eq.\,(\ref{eq:voltage_equation}) read {$V_{\mathrm Na}=50\,$mV},
{$V_{\mathrm K}=-77\,$mV}, {$V_{\mathrm L}=-54.4\,$mV}, and $C=1\,
\mu$F/cm$^2$. Furthermore,  the leakage conductance is assumed to be
constant, {$G_{\mathrm L} =0.3\,$mS/cm$^2$}. On the contrary, the
sodium and potassium conductances are controlled by the
voltage-dependent gating dynamics of single ion channels and are
proportional to their respective numbers. In the Hodgkin-Huxley model \cite{hh}, the opening of the
potassium ion channel is governed by four identical activation gates, being
characterized by the opening probability $n$. The channel is open
when all four gates are open. In the case of sodium channel, the
dynamics is governed by a set of  three independent and identical fast
activation gates ($m$) and an additional slow, so-termed
inactivation gate ($h$). The independence of the gates implies that
the probability $P_{\mathrm K, \mathrm Na}$ of the occurrence of the
conducting conformation is $P_{\mathrm K}=n^4$ for a potassium
channel and $P_{\mathrm Na}=m^3\, h$ for a sodium channel,
respectively. In a mean field description, the macroscopic
potassium and sodium conductances then read:
\begin{equation}
\label{eq:conductances-hodgkinhuxley}
G_{\mathrm{K}}(n)=g_{\mathrm{K}}^{\mathrm{max}}\ n^{4} , \quad
G_{\mathrm{Na}}(m,h)=g_{\mathrm{Na}}^{\mathrm{max}}\ m^{3} h\, ,
\end{equation}
where {$g_{\mathrm K}^{\mathrm max}=36\,$mS/cm$^2$} and
{$g_{\mathrm Na}^{\mathrm max}=120\,$mS/cm$^2$} denote
the maximal conductances
(when all channels are open). The
two-state, opening--closing dynamics of the gates
is governed by the voltage dependent opening and closing
rates $\alpha_x(V)$ and $\beta_x(V)\; (x=m,h,n)$, i.e., \cite{hh}
\begin{eqnarray}
\label{eq:rates}
\alpha_n(V)  =  \frac{0.01 \left(V+55 \right) }{1-\exp\left[-\left( V+55\right) / 10 \right] } \, ,  \\ \beta_n(V) = 0.125 \exp \left[ - \left( V + 65\right) / 80 \right] \, , \\
\alpha_m(V)  =  \frac{0.1  \left(V+40\right)}{1-\exp\left[-\left( V+40\right) / 10 \right] }\, , \\ \beta_m(V) = 4 \exp \left[ - \left( V+65 \right)/ 18\right] \, , \\
\alpha_h(V)  =  0.07 \exp\left[ -\left( V+ 65\right)/20\right] \, , \\ \beta_h(V) = \frac{1}{1+ \exp \left[ - \left( V + 35 \right) / 10\right]} \, .
\end{eqnarray}
Hence, the dynamics of the opening probabilities for the gates read:
\begin{equation}
  \label{eq:gating}
  \dot{x} = \alpha_{x}(V)\ (1-x)-\beta_{x}(V)\ x,\quad x=m,h,n\, .
\end{equation}
The voltage equation (\ref{eq:voltage_equation}), Eq.~(\ref{eq:conductances-hodgkinhuxley})
and the rate equations of the gating dynamics Eqs.~(\ref{eq:rates})- (\ref{eq:gating})
then constitute the original, strictly deterministic Hodgkin--Huxley model
 for spiking activity of the squid giant axon.

\subsection{Modelling Channel Noise}

In this study, however,  each channel defines a bistable stochastic element which
fluctuates between its {\itshape closed and open states}.
As a consequence, the number of open channels undergoes a birth-death stochastic process.
Applying a diffusion approximation to this discrete dynamics, the resulting
Fokker-Planck equation can be obtained from a Kramers-Moyal expansion \cite{tuckwell, noi1}.
The corresponding Langevin dynamics, interpreted here in the stochastic It$\hat{o}$ calculus \cite{PR1982},  reads:
\begin{equation}
 \label{eq:stochgating}
  \frac{\mathrm{d}}{\mathrm{d}t} x = \alpha_x(V) \, \left( 1 - x \right) - \beta_x(V)\, x   + \xi_x(t)\, , \quad x = n, m, h\, .
\end{equation}
It is driven by independent Gaussian white noise sources $\xi_x(t)$ of
vanishing mean which account for the fluctuations of the
number of open gates. The (multiplicative) noise strengths depend on
both, the membrane voltage and the gating variables. Explicitly,
these noise correlations assume the following form for a
neuron consisting of  $N_{\mathrm Na}$ sodium and
$N_{\mathrm K}$ potassium ion channels:
\begin{eqnarray}
 \langle \xi_m(t)\xi_m(t') \rangle &=& \frac{(1-m)\alpha_m+m \beta_m}{N_{Na}}\delta(t-t'),\\
  \langle \xi_h(t)\xi_h(t') \rangle &=& \frac{(1-h)\alpha_h+h\beta_h}{N_{Na}}\delta(t-t'),\\
\label{eq:noise_stregnth}
  \langle \xi_n(t)\xi_n(t') \rangle &=& \frac{(1-n)\alpha_n+n\beta_n}{N_K}\delta(t-t').
\end{eqnarray}

The fluctuations of the number of open ion channels result in conductances fluctuations of
the cell membrane eventually leading to spontaneous action potentials.
These spontaneous spiking events occur even for sub-threshold, constant external current stimuli,
i.e. for $I_{i, \mathrm{ext}}<I_1$. If the times of spike occurrences are given by $t_n$ with $n=0,1,2,...,N$, where $N$
is the number of observed spikes, the interspike interval between two succeeding
spike is $T_n = t_{n} - t_{n-1}$ ($n=1,.., N$). For the case of a single Hodgkin-Huxley neuron
the distribution of these interspike intervals exhibits a peak-like structure with the
peak located around the intrinsic time $T_\mathrm{intrinsic}$, which is determined by the limit
cycle of the deterministic dynamics \cite{Chow1996}.

The strength of the channel noise scales inversely with the number of the ion channels.
Consequently, the threshold for excitation can be reached more easily
with increasing the noise strength (i.e. smaller system size).
In order to characterize the spontaneous spiking, we introduce the mean interspike interval
\begin{equation}
 \label{eq:meaninterspikeinterval}
 \langle T \rangle := \lim_{N \to \infty} \, \frac{1}{N} \sum_{n=1}^{N}  T_n \, .
\end{equation}
With increasing number of ion channels, i.e. decreasing channel noise level,
the mean interspike interval increases exponentially for a vanishing
current stimulus $I_\mathrm{ext}=0$ \cite{Chow1996}.

In presence of a finite  positive constant current stimulus $I_\mathrm{ext}$, the mean interspike interval
is always smaller than that for the undriven case \cite{Schmid2007}.
Moreover, for supra-threshold driving, i.e. $I_\mathrm{ext}>I_2$,
noise-induced skipping of spikes is observed.
Accordingly, the channel noise does not only favor the generation of spikes,
but can as well suppress deterministic spiking.
For intermediate constant current driving,
i.e.  $I_1 \le I_\mathrm{ext} \le I_2$,
for which the Hodgkin-Huxley model exhibits a bi-stability
between a spiking and a non-spiking solution, channel noise results
in transitions between these two states.

\subsection{Numerical Methods}

Our numerical results  are obtained via the numerical integration of
the stochastic dynamical system given by Eqs.~(\ref{eq:voltage_equation})-(\ref{eq:stochgating}).
Particularly, we apply the {\itshape stochastic Euler-algorithm} in order to integrate the underlying stochastic dynamics \cite{kloeden}.
An integration step $\Delta t=0.01 \un{ms}$ has been  used in
the simulations; for the generation of the Gaussian distributed
random numbers, the Box-Muller algorithm \cite{bm} has been  employed.
The occurrence of a spiking event in the voltage signal $V_i(t)$ is obtained
by upward-crossing of a  detection threshold value $V_0=0$.
It turns out that this threshold can be varied over a wide range
with no influence on the resulting spike train dynamics.

To ensure that throughout all times the non-negative gating variables
take on values solely between 0 (all gates are closed) and 1 (all gates are
open), we implemented numerically  reflecting boundaries at 0
and 1. Throughout this work we assume  a constant ratio of the numbers
of potassium and sodium channels which results from constant ion channel densities.
For this work we have assumed channel densities of $20$
potassium channels and $60$  sodium channels per $\mu \mathrm{m}^{2}$.

In performing the numerics we initially prepare each neuron in the rest state voltage value $V_\mathrm{rest}$. By applying a short current pulse
on one of the two neurons, we initialize an action potential in this neuron which later on can be echoed by the
delay-coupling.

\section{Synchronization}
\label{sec:two-neuron-network}

In order to investigate the temporal correlation between the spiking statistics of the two neurons,
we apply the linearly interpolated, instantaneous time-dependent phase $\Phi_i(t)$
of a stochastic spiking process of neuron $i$ ($i=1,2$) \cite{Callenbach2002, Freund2003}; i.e., for $t \in [t_{i, n},t_{i, n+1} ]$
\begin{equation}
 \label{eq:phasedefinition}
 \Phi_{i}(t) =  2\pi\, n + 2\pi\, \frac{ t-t_{i, n} }{ t_{i, n+1}-t_{i, n} } \, , \quad n=0,1,2,...N_i-1\, ,
\end{equation}
where $t_{i, n}$ denotes the $n$-th spiking of neuron $i$ and $N_i$ is the total number of spike events in the
dynamics of neuron $i$. Note, that each spike occurrence contributes to the overall phase with $2\pi$.
Between two succeeding spikes the phase is obtained by linear interpolation.

Note, that in accordance with this definition of the phase,
the angular spiking frequency $\omega_i$ is given by:
\begin{equation}
 \label{eq:frequency}
 \omega_i = \lim_{t\to\infty} \frac{\Phi_i(t)}{t} \, .
\end{equation}
This frequency $\omega_i$ and the mean spiking rate being the inverse of
the mean interspike interval $\langle T \rangle_i$
in Eq.~(\ref{eq:meaninterspikeinterval}) are equivalent up to a constant factor; i.e. we obtain,
\begin{equation}
 \label{eq:freq_mii}
 \omega_i \doteq \frac{2 \pi}{\langle T \rangle_i}\, .
\end{equation}
Due to this equivalence we present our simulation results
solely in terms of the mean interspike interval $\langle T \rangle$.

In the pursuit  for a measure for the synchronization of spike occurrences in
the two coupled neurons, we consider the phase difference
between the two spike trains, namely:
\begin{equation}
 \label{eq:phasedifference}
 \Delta \Phi (t) = \Big( \Phi_1(t) - \Phi_2(t)\Big )\;\; \mathrm{mod} \, 2\pi \;.
\end{equation}
It turns out that the phase difference $\Delta \Phi(t)$ numerically tends, after transient effects have faded away, for $t\to \infty$
to a value  that depends on various parameters such as the coupling strength, the delay time, the total integration time and to some extent (due to inherent irregular, chaotic behavior) even on the time step used in our integration step. Accordingly, we observe in the long time limit that
$\omega_1=\omega_2$ and $\langle T \rangle_1 = \langle T \rangle_2 =: \langle T \rangle$.

\subsection{Deterministic limit}
\label{sec:two-neuron-network-det}

We  consider first the deterministic limit by  letting $N_\chem{Na} \to \infty$ and $N_\chem{K} \to \infty$.
In doing so, the occurrence of repetitive firing was systematically analyzed
by varying the two coupling parameters, i.e. the coupling strength $\kappa$ and the delay time $\tau$.

\begin{figure}
 \centerline{\includegraphics[width=1.0\linewidth]{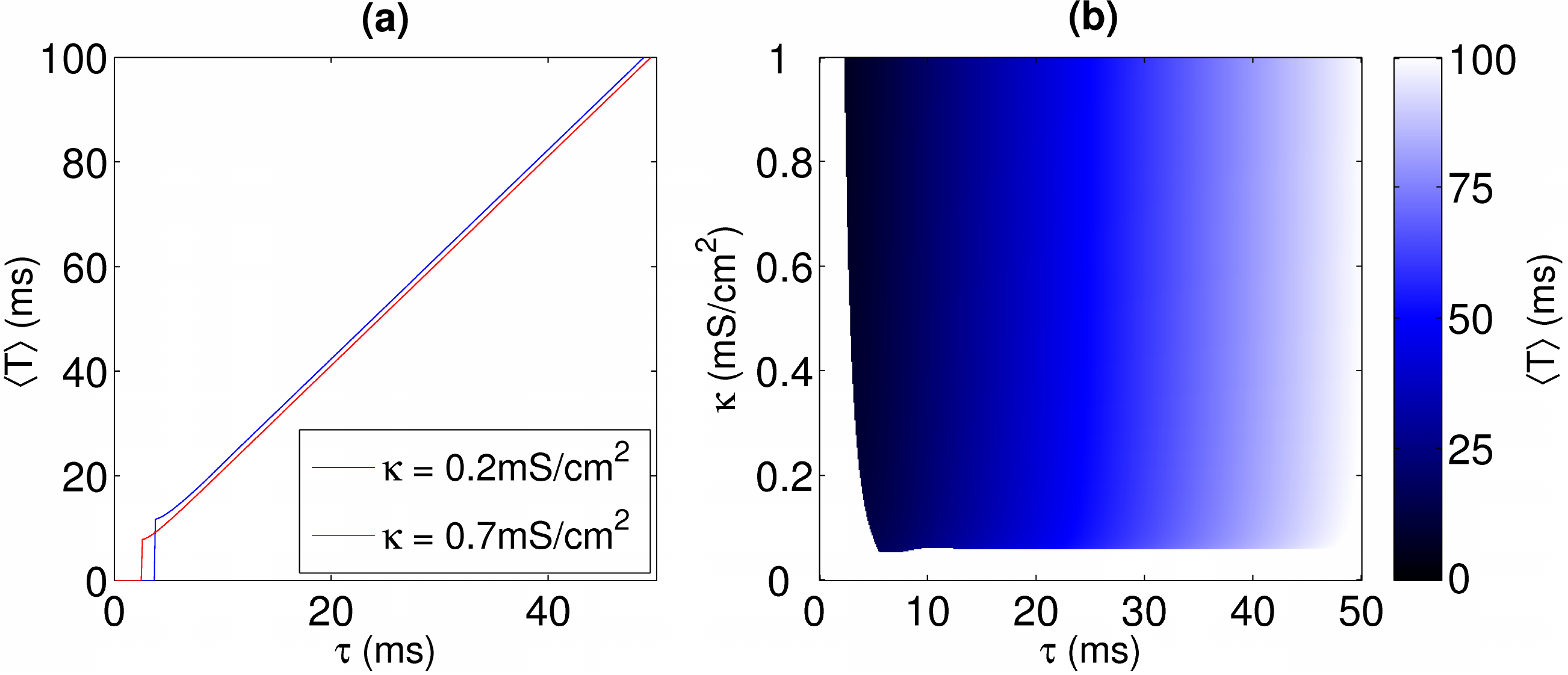}}
 \caption{ \label{fig:mii_det_i0}(color online)
  The interspike interval $\langle T \rangle$ for the spiking of one of
  the two delay-coupled standard Hodgkin-Huxley neurons with no external constant current stimulus,
  i.e. $I_\mathrm{ext}=0$, as function of the delay time $\tau$ for coupling strength $\kappa=0.2\un{mS/cm^2}$ in panel (a) and as function of both parameters of the delay-coupling, i.e. the coupling strength $\kappa$ and the delay time $\tau$, in panel (b). The color encoding for the phase diagram in panel (b) is shown in the color bar: the darker the color the shorter is the interspike interval $\langle T \rangle$.
 For white regions, the mean interspike interval $\langle T \rangle$ tends to infinity and no repitive firing is observed.  }
\end{figure}

For $I_\mathrm{ext}=0$, i.e. for a subthreshold constant current driving,
the resulting mean interspike interval $\langle T \rangle$ is depicted in Fig.~\ref{fig:mii_det_i0}(a).
For  the coupling parameters taken within the white region of Fig.~\ref{fig:mii_det_i0}(b),
the system relaxes to the non-spiking rest state.
However, in the regime of the spiking dynamics  a spike in any of the two neurons generates a subsequent spike   in the other neuron.
Consequently, repetitive, but alternating, firing can be observed for both neurons.
The mean interspike interval $\langle T \rangle$
increases linearly with increasing delay time $\tau$,
cf. Fig.~\ref{fig:mii_det_i0}(a). In particular,
\begin{equation}
\label{eq:detmisic}
\langle T \rangle \approx 2 ( T_\mathrm{act} + \tau )\, ,
\end{equation}
where $T_\mathrm{act} \approx 2 \mathrm{ms}$ is the activiation time which is the time
between the time the stimulus of delayed coupling sets in and
the occurrence of the stimulated spiking.
The factor \lq$2$\rq\, in Eq.~(\ref{eq:detmisic}) is due to  the  alternating spiking of the two neurons. Note, that
the time between succeeding spiking events of the network is the delay time plus the activation time.

\begin{figure}
  \centerline{\includegraphics[width=1.0\linewidth]{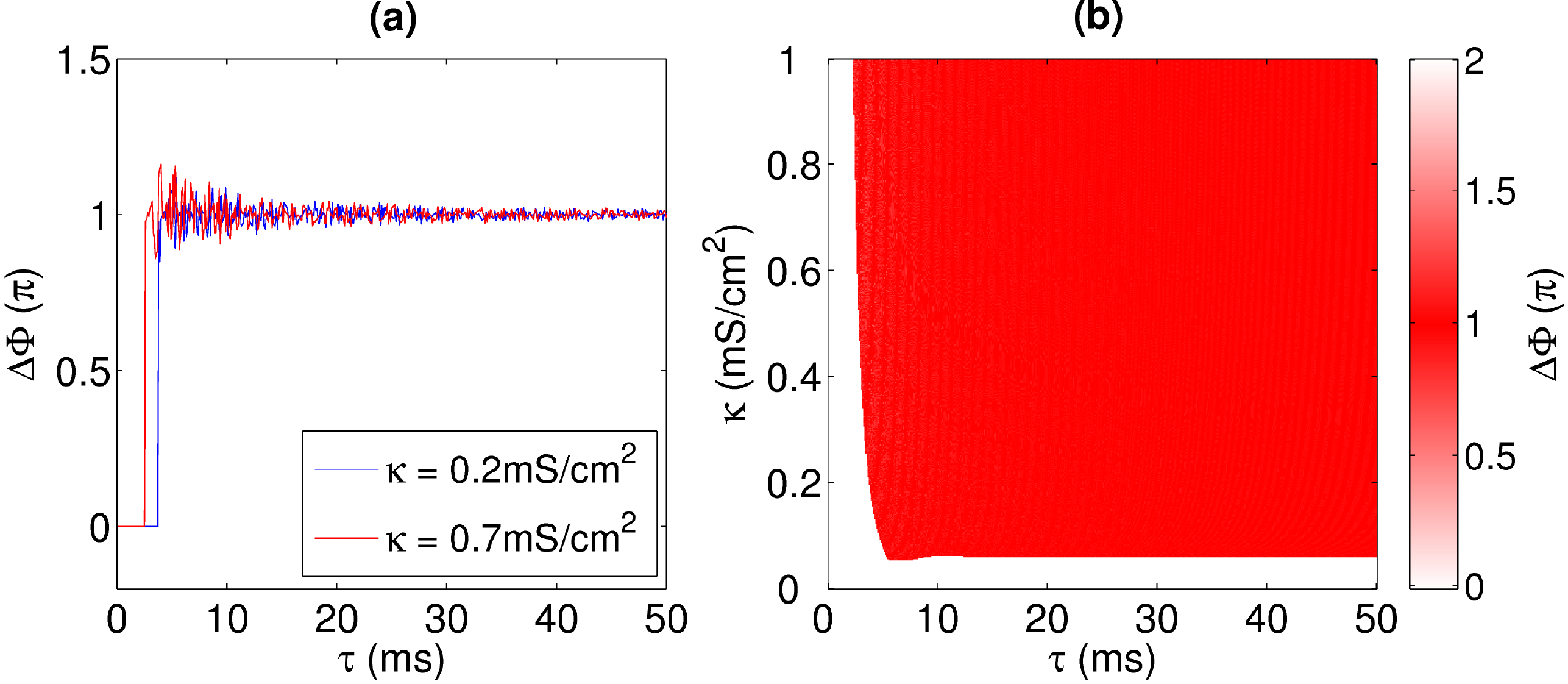}}
  \caption{ \label{fig:ph_det_i0}(color online)
  The phase difference $\Delta \Phi$, cf.  Eq.~(\ref{eq:phasedifference}),
  for two delay-coupled standard Hodgkin-Huxley neurons for $I_\mathrm{ext}=0$,
  cf. Eqs.~(\ref{eq:voltage_equation})-(\ref{eq:gating}), is depicted as function
  of the delay time $\tau$ in panel (a). In panel (b) the phase difference $\Delta \Phi$
  is depicted within a phase diagram upon varying the two coupling parameters $\kappa$ and $\tau$.
  In the white region, there is no repetitive firing, cf. Fig.~\ref{fig:mii_det_i0}.
  In the red region the two neurons fire alternatingly, resulting in a phase difference of $\Delta \Phi \simeq \pi$.}
\end{figure}

In Fig.~\ref{fig:ph_det_i0} the steady-state phase difference $\Delta \Phi$ is
depicted as function of the coupling strength  $\kappa$ and the delay time $\tau$.
For large delay times $\tau$ the neuronal dynamics of an individual neuron possesses after each spiking event  sufficient time
to relax back to its rest  state before the delayed stimulus caused by the spiking event of the other neuron sets in.
This results in an alternating firing dynamics of the two neurons. The spiking of the two neurons therefore exhibits a constant
phase shift of $\pi$, i.e. the spiking event of one neuron is almost perfectly located between two succeeding spiking events in the other neuron.
However, for delay times that are of the order, or are even smaller than the refractory time, an irregular behavior of the Hodgkin-Huxley
dynamics emerges, as it is to be expected in presence of  finite delay. This in turn is reflected in the numerically evaluated  phase difference $\Delta \Phi$ by a noisy behavior that succeedingly smoothes for increasing large delay times, yielding the afore mentioned asymptotic phase shift of $\pi$. This result is corroborated with our numerics,  as depicted with Fig.~\ref{fig:ph_det_i0}.

\subsection{Influence of channel noise}
\label{sec:channel_noise_network}

\begin{figure}
  \centerline{\includegraphics[width=1.0\linewidth]{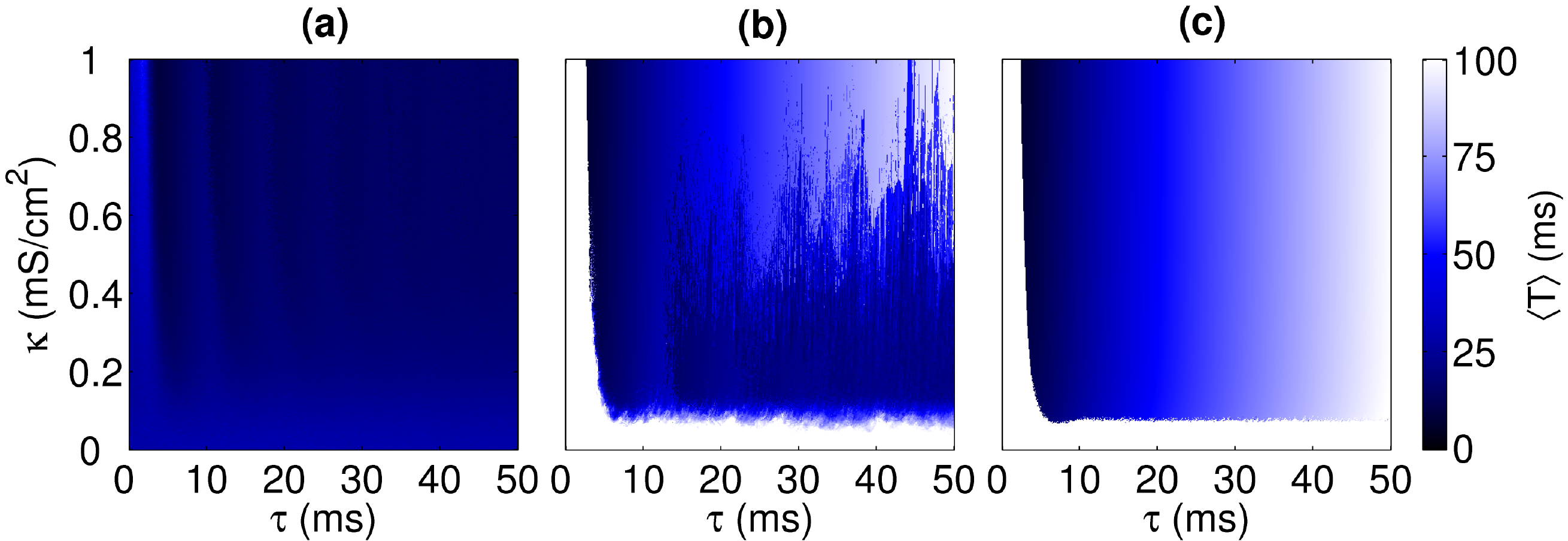}}
 \caption{ \label{fig:mii_st_i0} (color online)
  The interspike interval $\langle T \rangle$ for the spiking of one of
  the two delay-coupled standard Hodgkin-Huxley neurons with no external constant current stimulus,
  i.e. $I_\mathrm{ext}=0$, as function of the coupling parameters $\kappa$ and the delay time $\tau$.
  The mean interspike interval $\langle T \rangle$ is depicted for three different strengths of the
  channel noise: (a) strong intrinsic channel noise with $N_\chem{Na}=360\, ,\quad N_\chem{K}=120$,
  (b) moderate intrinsic channel noise with $N_\chem{Na}=3600\, ,\quad N_\chem{K}=1200$ and
  (c) weak intrinsic channel noise with $N_\chem{Na}=36000\, ,\quad N_\chem{K}=12000$.   }
\end{figure}

\begin{figure}
  \centerline{\includegraphics[width=1.0\linewidth]{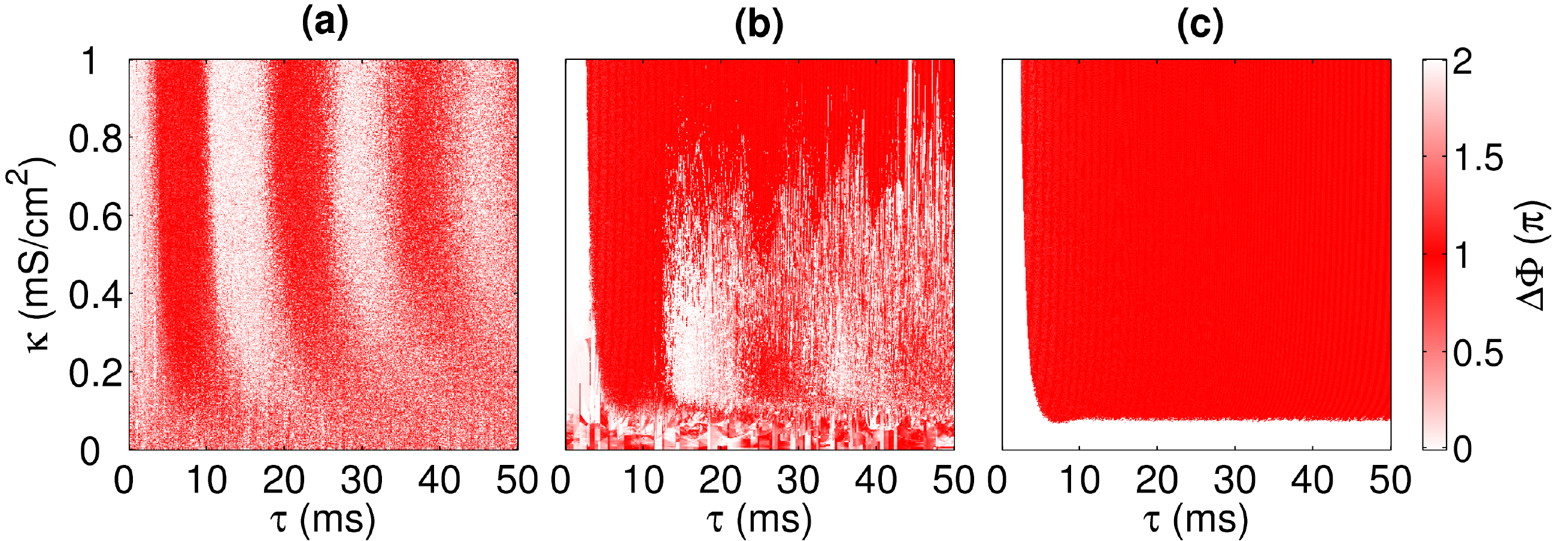}}
  \caption{ \label{fig:ph_st_i0}(color online) The steady-state phase difference $\Delta \Phi$
  for two delay-coupled stochastic Hodgkin-Huxley neurons for $I_\mathrm{ext}=0$ is depicted
  as function of the delay coupling parameters $\kappa$ and $\tau$ for three different
  noise strengths:  (a) strong intrinsic channel noise with $N_\chem{Na}=360\, ,\quad N_\chem{K}=120$,
  (b) moderate intrinsic channel noise with $N_\chem{Na}=3600\, ,\quad N_\chem{K}=1200$ and
 (c) weak intrinsic channel noise with $N_\chem{Na}=36000\, ,\quad N_\chem{K}=12000$.
  The stripes in panel (a) indicate noise-induced phase-flips, see text.}
\end{figure}

In Fig.~\ref{fig:mii_st_i0} we depict the dependence of the mean interspike interval $\langle T \rangle$
on the coupling parameter $\kappa$ and  delay time $\tau$ for three different levels of channel noise,
i.e. different sets of ion channel numbers $N_\chem{Na}$ and $N_\chem{K}$.
With increasing noise level, i.e. decreasing number of ion channels, the sharp transition between
the parameter regime of repetitive spiking and non-repetitive spiking is smeared out.

\begin{figure}
 \centerline{\includegraphics[width=1.0\linewidth]{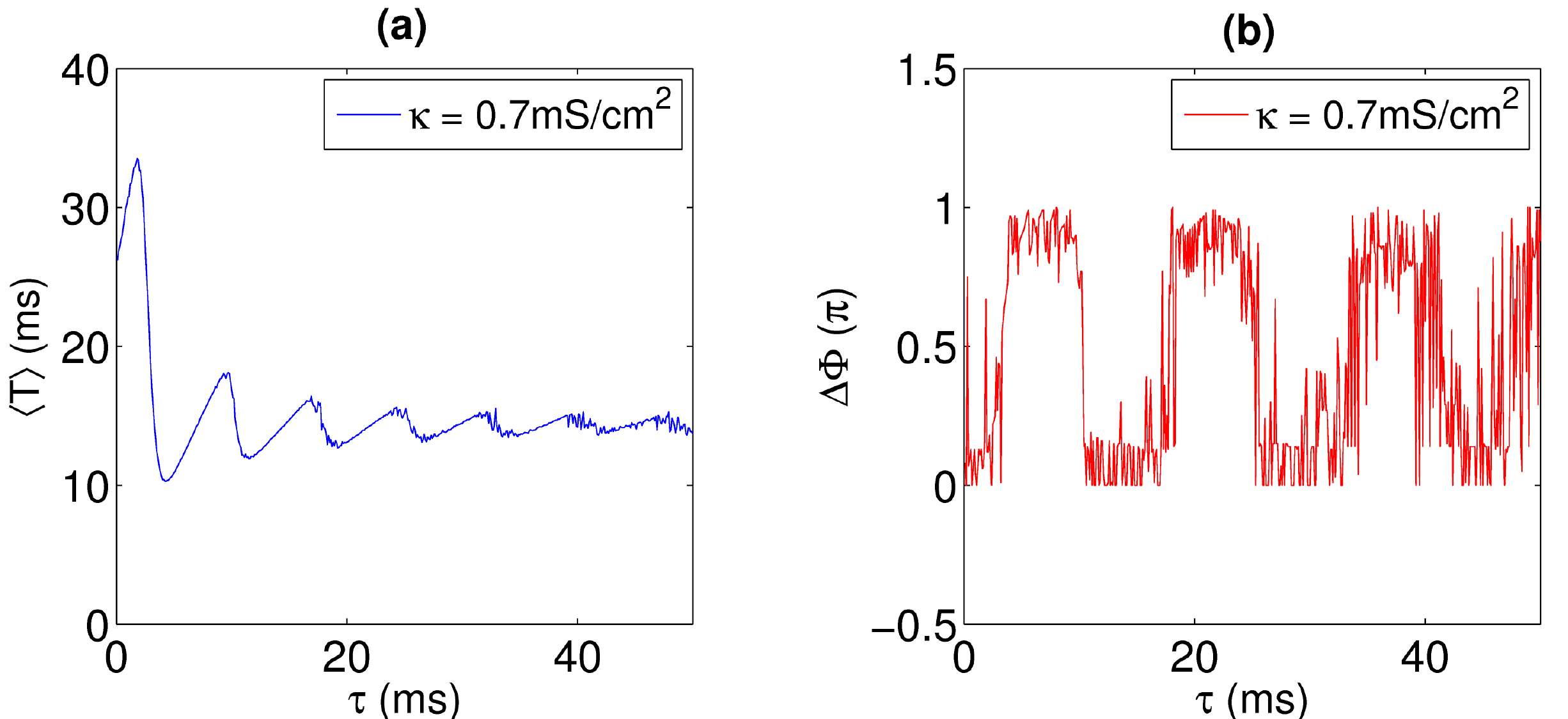}}
 \caption{\label{fig:phaseflip}(color online)
 In panel (a) the mean interspike interval $\langle T \rangle$ is depicted as function of the
  coupling time $\tau$ for constant coupling strength $\kappa$,
  $I_\mathrm{ext}=0$ and channel noise level corresponding to
  numbers of ion channel: $N_\chem{Na} = 360$ and $N_\chem{K}=120$.
  For the same parameters, the phase difference $\Delta \Phi$
  as function of the delay time
  $\tau$
  is depicted in (b).}
\end{figure}

In addition, for considerable {\it strong} channel noise, distinct synchronization patterns emerge, indicating a frequency locking
similar to the one observed for the autaptic case discussed in Sec.~\ref{sec:retrospect} below. In order to analyze the observed synchronization patterns in greater detail, we depict the mean interspike interval
$\langle T \rangle$ as function of the delay time $\tau$ for a fixed coupling strength $\kappa=0.7 \un{mS/cm^2}$ in Fig.~\ref{fig:phaseflip}(a). We find that the mean interspike interval varies with the delay time in an almost piecewise linear manner, displaying sharp, triangle-like
textures.

The distinct peak locations of the mean interspike interval $\langle T \rangle$ can be explained
by the  number of spikes that match in accordance with the intrinsic time $T_\mathrm{intrinsic}$
a full propagation time length,
given by twice the delay time. The mean interspike interval $\langle T \rangle$
henceforth is proportional to the ratio of twice the delay time and the
number of spikes fitting into this very delay time interval, cf. Eq.~(\ref{eq:detmisic}).
Accordingly, channel noise results in a {\it stabilization} of the interspike interval to an interval around
the intrinsic time scale of the Hodgkin-Huxley oscillator.

\subsection{Phase-flip bifurcation}
\label{sec:phase-flip}

\begin{figure}
 \centerline{\includegraphics[width=1.0\linewidth]{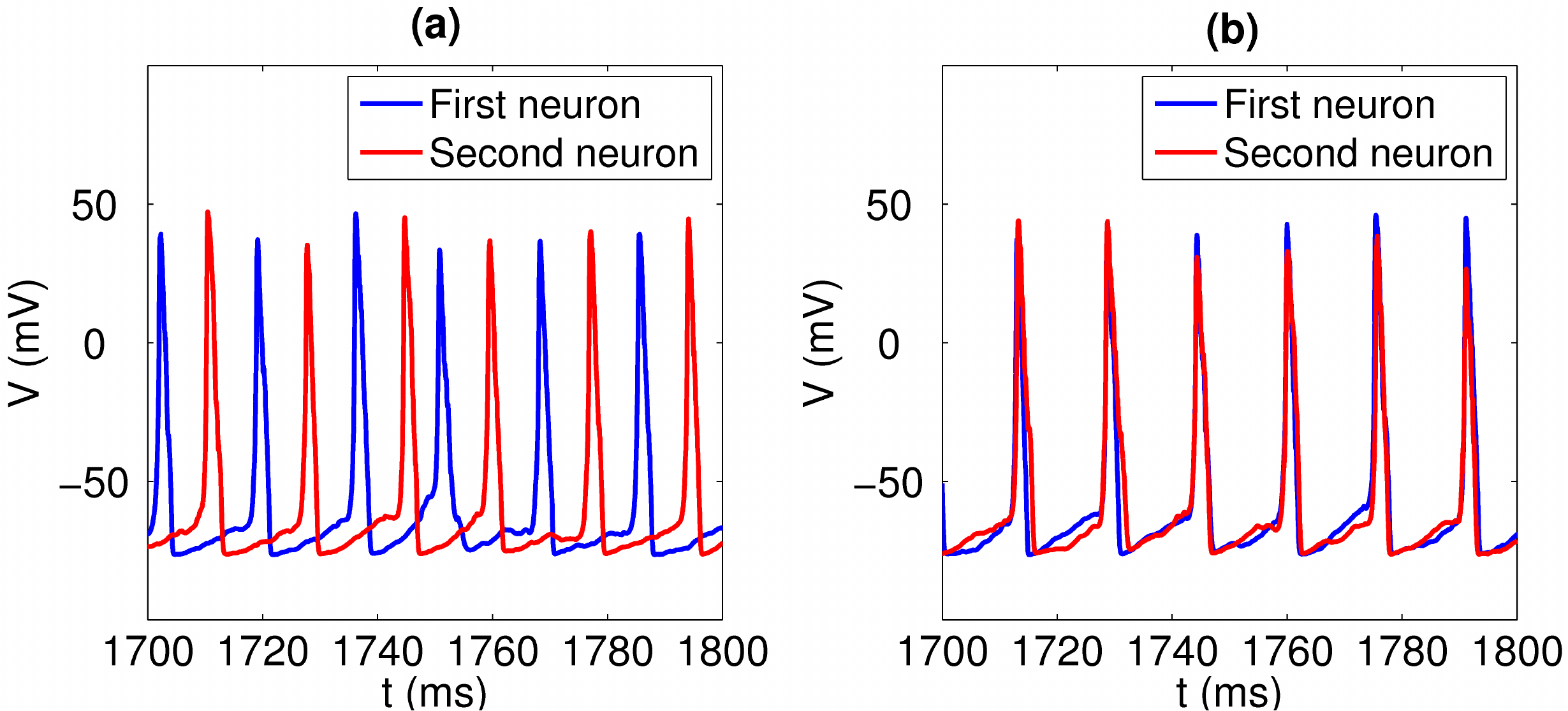}}
 \caption{\label{fig:spiketrains}(color online) The simulated time-course of the membrane potentials for the two delayed coupled Hodgkin-Huxley neurons is depicted:
 (a) for the parameters $\kappa= 0.7{\mathrm{mS/cm}}^2$ and $\tau=8\mathrm{ms}$,the dynamics shows alternating, i.e. {\itshape antiphase},
 spiking, (b) for $\kappa= 0.7\mathrm{mS/cm}^2$ and $\tau=15\mathrm{ms}$ synchronous, {\itshape inphase} firing of the
 two neurons is observed. }
\end{figure}

Note, that this phenomenon complies with similar pattern appearing in the diagram for the phase difference,
cf. Fig.~\ref{fig:ph_st_i0}. Furthermore,
we depict the phase difference $\Delta \Phi$ in Fig.~\ref{fig:phaseflip}(b). These sharp transitions are accompanied by pronounced noise-induced phase-flips. At the corresponding phase-flip values at $0$ and $\pi$ the spiking of the basic network changes from an in-phase towards an anti-phase spiking: For $\Delta \Phi \approx 0$, both neurons spike simultaneously, implying  synchronous firing, cf. Fig.~\ref{fig:spiketrains}.

\section{Retrospect: Frequency locking by an autaptic feedback loop}
\label{sec:retrospect}

The observed frequency stabilization in the dynamics of a network of two delayed coupled neurons shares it's origin with the frequency locking phenomenon in noisy neurons
with an autaptic feedback-loop. When neuronal dendrites establish an autapse, i.e. a connection to the neuron's own axon, a delayed feedback loop
is induced to the neuron's dynamics. Such autosynapses,  described originally by Van der Loos and Glaser in 1972
\cite{Loos1972} are a  common phenomenon found in about $80\%$ of all analyzed pyramidal cells in the
cerebral neocortex of human brain \cite{Lubke1996}.

Auto-synapses establish a time-delayed feedback mechanism on the cellular level \cite{Loos1972}.
From a mathematical point of view, autaptic connections introduce new time scales
into the single neuron dynamics which gives raise to peculiar frequency looking behavior \cite{Li2010}.
Our modelling, cf. Eqs.~(\ref{eq:voltage_equation})-(\ref{eq:stochgating}),
captures the stochastic autapse phenomena for $i=j$ and only one neuron.

\begin{figure}
 \centerline{\includegraphics[width=1.0\linewidth]{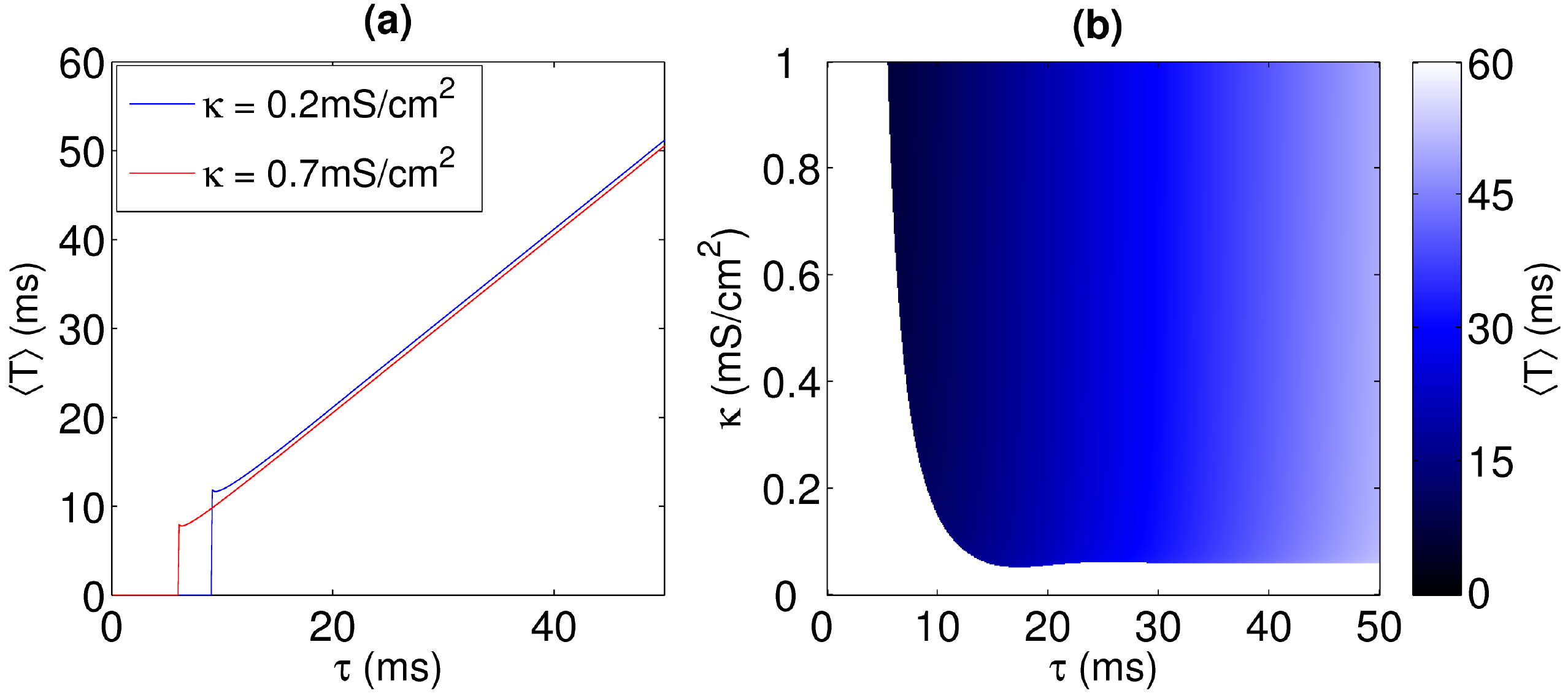}}
 \caption{ \label{fig:autapse}(color online)
 The interspike interval $\langle T \rangle$ for the standard Hodgkin-Huxley
 neuron with an autaptic feedback loop, cf. Eqs.~(\ref{eq:voltage_equation})-(\ref{eq:gating}).
 In panel (a) the dependence of the interspike interval $\langle T \rangle$ on the delay time $\tau$
 is depicted for different coupling strengths $\kappa$.
 In the regime of repetitive firing the mean interspike interval $\langle T \rangle$
 grows linearly with the delay time $\tau$.
 In the phase diagram shown in panel (b) the dependence of the interspike interval $\langle T \rangle$
 on the coupling strength $\kappa$ and the delay time $\tau$ is depicted.
 The color bar next to panel (b) gives the color encoding for the values of the interspike intervals.
 The white region corresponds to the situation for which the externally initialized spike is not
 echoing itself and, formally, $\langle T \rangle \to \infty$.}
\end{figure}

In the limit of vanishing channel noise by letting
$N_\chem{Na} \to \infty$ and $N_\chem{K} \to \infty$,
the spiking period is given by the delay time $\tau$ plus the activation time $T_\mathrm{act}$, being
the time needed for creating the next spike event after the delayed stimulus did set in, cf. Fig.~\ref{fig:autapse}(a).
Note, that in presence of the autaptic delay, the fixed-point solution of the unperturbed
Hodgkin-Huxley dynamics remains stable and the delayed stimulus
is excitatory.

\begin{figure}
 \centerline{\includegraphics[width=1.0\linewidth]{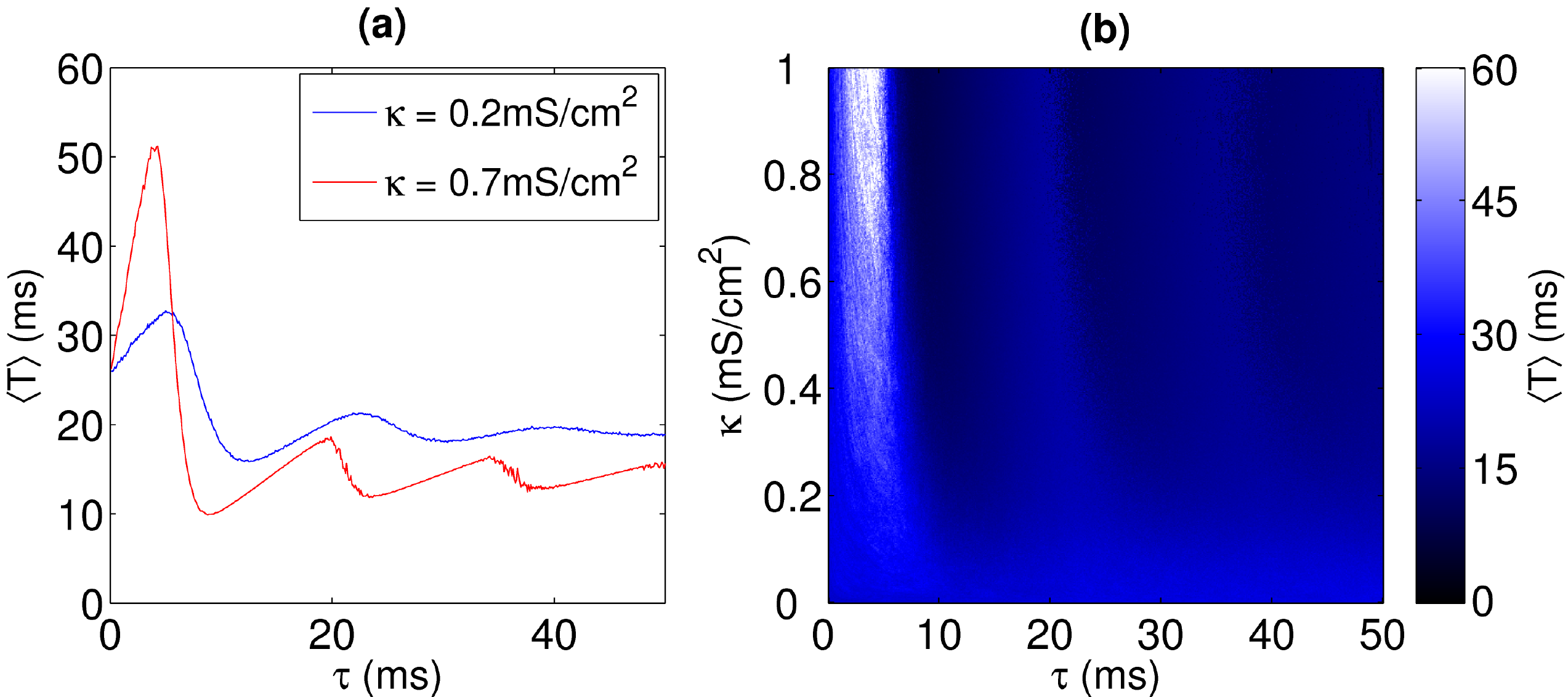}}
  \caption{ \label{fig:stochastic_autapse}(color online)
  The mean interspike interval $\langle T \rangle$ for the stochastic
  Hodgkin-Huxley neuron with an autaptic feedback loop, cf. Eqs.~(\ref{eq:voltage_equation})-(\ref{eq:noise_stregnth}).
  The channel noise level corresponds to $N_\chem{Na} = 360$ and $N_\chem{K}=120$.
  Similarly to Fig.~\ref{fig:autapse}, $\langle T \rangle$ is shown as function of
  the delay time $\tau$ in panel (a), and the corresponding
  phase diagram is depicted in panel (b). The color bar gives the color
  encoding of the mean interspike intervals $\langle T \rangle$.}
\end{figure}

In presence of finite channel noise, however, there are two competitive mechanisms at work:
(i) there are spiking repetitions due to the delay-coupling and
(ii) noise-induced generation of spikes or (iii) noise-induced skipping of spikes.
The interplay between these mechanisms becomes evident when the
distribution of the interspike intervals is considered.
In particular, the interspike interval histogram (ISIH) exhibits a bimodal
structure, exhibiting two peaks, see in Fig.~6 in Ref. \cite{Li2010}.
Upon  specific values of the noise strength determined by the number of ion channels
and the coupling parameters $\kappa$ and $\tau$
the bimodal structure shows distinct differences: Due to first mechanism the delay coupling
leads to a significant peak around the delay time $\tau$. Via the noise-induced mechanisms, the channel noise
results in a broad peak around the intrinsic time scale $T_\mathrm{intrinsic}$.

However, for considerable strong coupling strengths the bimodal structure
collapses to an unimodal one and a frequency-locking phenomena takes place.
The mean interspike interval $\langle T \rangle$ becomes bounded by a finite range of values around the intrinsic time scale,
but still shows
a striking dependence on the delay time $\tau$, cf. Fig~\ref{fig:stochastic_autapse}.  In particular, the mean interspike interval $\langle T \rangle$
varies with the delay time $\tau$ in an
almost piecewise linear manner, displaying sharp triangle-like
textures, cf. Fig.~\ref{fig:stochastic_autapse}(a)
for $\kappa=0.7\un{mS/cm^2}$ and $\tau>10\un{ms}$.

\section{Conclusions}
\label{sec:conclusion}

With this work we have investigated the effects of intrinsic channel
noise  on the spiking dynamics of an elementary  neuronal network
consisting of two {\it stochastic} Hodgkin-Huxley neurons.
In doing so, we invoked some idealistic simplifications such as the use of
bilinear coupling with identical time delays and equal  coupling strengths.
The finite transmission time of an action potential traveling along the neuronal axon
to the dendrites is the cause for the  delay-coupling to the other neuron.
Physically this transmission time derives from the  finite length to the connecting dendrites
and the finite propagation speed.

A Pyragas-like delayed feedback mechanism has been employed to model the delayed coupling.
The two basic parameters for the delay-coupling are the coupling strength $\kappa$ and
the delay time $\tau$. In terms of  these two parameters the delayed feedback mechanism
results in a periodic, repetitive firing  events of the neurons.

Apart from this repetitive firing, the delay-coupling introduces intriguing time scales.
Particularly, we detect a noise-induced locking of the spiking rate to the
intrinsic frequency of the system. Consequently, the delayed feedback mechanism
serves as a control option for adjusting
the interspike intervals; this feature may be of importance for
memory storage \cite{Seung2000} and stimulus locked short-term dynamics in
neuronal systems \cite{Tass2002}. One may therefore speculate whether ubiquitous intrinsic channel
noise in combination with the autapse phenomenon is constructively harvested by
nature for efficient frequency filtering.

Moreover, the emergence of  a correlation between the firing dynamics of the two delayed
coupled neurons has been studied in terms of {\it stochastic} phase synchronization.
The dynamics of two coupled Hodgkin-Huxley neurons exhibits noise-induced phase-flip bifurcations.
At these phase-flip bifurcations the  phase difference changes abruptly and
the spiking of the neuron switches from an in-phase spiking to
an anti-phase spiking, or vice versa.
These phase-flips are the direct result  of the influence of channel noise.
The observed phase-flips thus may possibly assist the fact of a coexistence of various
frequency rhythms and oscillation patterns in different parts of extended
neuronal networks, as for example it is the case for the brain.

\section*{Acknowledgments}

This work was supported  by the Volkswagen foundation
project I/83902 (PH, GS) and by the German excellence
cluster ``Nanosystems Initiative Munich II'' (NIM II) via
the seed funding project (XA, GS, PH).

\section*{References}

\bibliographystyle{apsrev4-1}
%

\end{document}